\documentclass[apj,numberedappendix]{emulateapj}
\usepackage{pxfonts}
\usepackage{color}

\shorttitle{Gamma-ray Emissions from Low-Mass X-ray Binaries}
\shortauthors{Zhang et al.}

\begin{document}

\title{Potential Gamma-ray Emissions from Low-Mass X-ray Binary Jets}

\author{Jian-Fu Zhang\altaffilmark{1,2}, Wei-Min Gu\altaffilmark{1}, Tong Liu \altaffilmark{1}, Li Xue\altaffilmark{1} and  Ju-Fu Lu\altaffilmark{1}}

\altaffiltext{1}{Department of Astronomy and Institute of Theoretical Physics and Astrophysics, Xiamen University, Xiamen, Fujian 361005, China;}
 \email{jianfuzhang.yn@gmail.com(JFZ); guwm@xmu.edu.cn(WMG); lujf@xmu.edu.cn(JFL)}  %
\altaffiltext{2}{Department of Physics, Tongren University, Tongren, Guizhou, 554300, China}

\begin{abstract}
By proposing a pure leptonic radiation model, we study the potential gamma-ray emissions from jets of the low-mass X-ray binaries. In this model, the relativistic electrons that are accelerated in the jets are responsible for radiative outputs. Nevertheless, dynamics of jets are dominated by the magnetic and proton-matter kinetic energies. The model involves all kinds of related radiative processes and considers the evolution of relativistic electrons along the jet by  numerically solving the kinetic equation. Numerical results show that the spectral energy distributions can extend up to TeV bands, in which synchrotron radiation and synchrotron self-Compton scattering are dominant components. As an example, we apply the model to the low-mass X-ray binary GX 339--4. The results can not only reproduce the currently available observations from GX 339--4, but also predict detectable radiation at GeV and TeV bands by \emph{Fermi} and CTA telescopes. The future observations with \emph{Fermi} and CTA can be used to test our model, which could be employed to distinguish the origin of X-ray emissions.
\end{abstract}


\section{Introduction}
\label{Intro}
Among Galactic X-ray binaries, there are about twenty strong candidate or firmly confirmed microquasars that present the extended relativistic radio jets, in which the jet is powered by an accretion of the central compact object. According to the mass of the companion star, they are classified as the high- and low-mass microquasars. For this kind of X-ray binaries, it is widely believed that during the low/hard spectral state, radio and infrared (IR) emissions originate from synchrotron processes of relativistic electrons in a persistent jet.

In the low/hard state of these X-ray binaries, there is still ongoing debate regarding the origin of X-ray emissions \citep[e.g.,][]{McClintock06}. One possibility is that they originate in a hot plasma corona \citep[e.g.,][]{Liu99}, through the Comptonization of the thin accretion disk photons by hot electrons in the corona, or in a hot accretion flow \citep[e.g.,][]{Esin01,Yuan05}, via the Comptonization of synchrotron photons within the hot accretion flow. Another possibility is that they may have their origins from the jets, either as synchrotron emission \citep[e.g.,][]{Markoff01,Markoff03,Kaiser06,Peer09,Maitra09,Vila10,Peer12,Vila12} or synchrotron self-Compton scattering \citep[SSC,][]{Markoff05,Maitra09}, or external-radiation-Compton scattering \citep[ERC,][]{Peer12}. Alternatively, the jet emitting disk \citep[e.g.,][]{Zhang13} and disk corona-jet models \citep[][]{Qiao15} are also proposed to interpret X-ray emissions from black-hole X-ray binaries.

The study of observational and theoretical aspects of microquasars, particularly in the GeV and TeV domains, is a very active topic. From an observational point of view, five sources (including confirmed and candidate microquasars) are detected at high-energy (GeV) bands by the \emph{Fermi} Large Area Telescope (\emph{Fermi} LAT) and the AGILE satellite, and/or at very high-energy (TeV) bands by the Major Atmospheric Gamma Imaging Cherenkov Telescope (MAGIC), the High Energy Stereoscopic System (HESS) and the Very Energetic Radiation Imaging Telescope Array System (VERITAS): Cygnus X-1 \citep[e.g.,][]{Albert07,Malyshev13}, Cygnus X-3 \citep[e.g.,][]{Abdo09a,Tavani09}, SS433 \cite[][]{Bordas15}, LS 5039 \citep[e.g.,][]{Aharonian05,Abdo09b}, and LS I +$61^{\circ}$ 303 \citep[e.g.,][]{Albert06,Abdo09c}. These gamma-ray detections imply that extreme particle acceleration is at work in these systems. However, none of these detected gamma-ray sources is a low-mass microquasar. One possible reason is that these systems cannot exactly produce any GeV and TeV photons, as a result of the expected inefficient effect of a low-mass companion star. Another possible reason is that one has not put forward an observational plan to detect this class of object. However, they are emitting  detectable gamma-ray signature. To our knowledge, almost no campaign detecting gamma-rays from a low-mass microquasar is carried out \citep[however, see][]{Bodaghee13}.

Based on leptonic and/or hadronic considerations, many models for jet emissions of high-mass microquasars are proposed in the literature \citep[e.g.,][]{Romero03,Bosch06,Khangulyan08,ZAA14,Zhang14}. In these models, the stellar companion plays a key role, providing a seed photon field in a leptonic model or a target for proton--proton/proton--photon interactions in a hadronic one, even resulting in a cascade process. However, very little attention is focused on the low-mass microquasars, in particular, for gamma-ray productions in a leptonic model. The hadronic model for a low-mass microquasar is proposed in \citet{Romero08}, in which theoretical spectral energy distributions (SEDs) from cascade processes induced by relativistic protons, can well extend into PeV bands. Subsequently, the further refined models (called as lepto-hadronic ones) are developed to investigate the low-mass microquasars GX 339--4 \citep[][]{Vila10} and XTE J1118+480 \citep[][]{Vila12}. In these lepto-hadronic models, radio and X-ray emissions are explained as an origin of the jet, in which SEDs could predict the detectable emissions at GeV and TeV bands by \emph{Fermi} LAT and CTA.

A recent search on neutrino emission from a sample of six microquasars is presented in \citet{Adri14} by the ANTARES neutrino telescope. This study has provided upper limits of the possible neutrino fluxes, which constrains the scenario that the ratio of proton to electron power is far less than 100 for a hadronic jet model of low-mass microquasars \citep{Zhang10}. Besides, almost all the fits to GX 339--4 in the lepto-hadronic models need the ratio of proton to electron power close to 1 \citep[e.g.,][]{Vila10}. In a hadronic or lepto-hadronic model, protons have to be accelerated to about 10 PeV in order to excite hadronic cascade emissions. Such a high energy actually requires the existence of an extreme physical environment. Nevertheless, the requirement for electrons in a hadronic model is relatively loose, that is, the maximum energy of electrons is commonly below TeV energies \citep[e.g.,][]{Romero08,Zhang10}. However, the simulations by particle-in-cell (PIC) methods demonstrate the presence of very strong coupling between electrons and shock-heated protons \citep[][]{Spitkovsky08,Sironi11}, which implies that the dissipated energies are mostly converted into primary electrons. Furthermore, the recent hybrid PIC simulations indicate that it needs a stringent condition for an effective acceleration of protons, that is, only when shocks propagate parallel to the upstream magnetic field \citep{Caprioli14}.

Given the above analysis and a lower radiative efficiency of protons than that of electrons, the leptonic models appear to be strongly supported over the hadronic ones. Therefore, in this work we carry out a study about broadband emissions of low-mass microquasars in the framework of a pure leptonic model. We want to know whether a pure leptonic model can also provide the expected gamma-ray emissions as a hadronic one. It should be noted that there are some works on the jet emission of X-ray binaries in the framework of the leptonic model \citep[][]{Kaiser06,Peer09}. A common feature is that they considered the effects of radiative and adiabatic cooling of relativistic electrons. More concretely, \citet{Kaiser06} focused on the flat synchrotron spectra of partially self-absorbed jets by considering the different jet geometries, such as ballistic and adiabatic scenarios. By using the electron spectral distribution with a Maxwellian shape at low energies plus a power-law tail at high energies, \citet{Peer09} studied the features of radiative spectra from the jet of X-ray binaries in detail and considered that a single acceleration episode is at work at the base of the jet. In these works, the models can produce multi-wavelength emissions ranging from radio to hard X-ray (or soft gamma-ray) bands and do not need to re-accelerate the electrons in the jet.

In particular, except for including the radiative and adiabatic losses of relativistic electrons, we consider the evolution of the relativistic electron along the jet by the kinetic equation. Furthermore, the dissipation region is extended to a large scale of the jet where the maximum energy of ultra-relativistic electrons can be constrained by the first-order \emph{Fermi} acceleration mechanism. It is more obvious that the SEDs in our work can extend into the GeV and TeV ranges. As an example of application, we use our model to fit the low-mass microquasar GX 339-4. The results show that our pure leptonic model can not only provide possible explanations for the current observations, but also predict GeV and TeV band emissions.

The paper is organized as follows. Descriptions for a leptonic jet model are presented in Section~2. Numerical results of the model are shown in Section~3. Section~4 is the fitting results to GX 339--4. Conclusions and discussion are given in Section~5.

\section{Model Descriptions}
\label{model}
Like the case of active galactic nuclei, the mechanism of jet formation of X-ray binaries remains an open question. The seminal theoretical works suggested that large-scale magnetic fields anchored in an accretion disk or a black hole induce the generation of jets by means of a magneto-centrifugal mechanism \citep{BZ77,BP82}. According to this mainstream thought,  the jet in this work is considered as a Poynting flux-dominated one in its innermost region and then converts the magnetic energy into the matter kinetic energy at larger jet height. However, it is unclear that for a Galactic jet, where and how it converts from the Poynting flux- to matter-dominated jets and how efficient the conversion is. It is stated that MHD instabilities are possible trigger mechanisms of the conversion \citep[e.g.,][]{Sikora05}. These instabilities could produce the magnetic energy dissipation via shocks or magnetic reconnection, resulting in an acceleration of relativistic particles.

In this study, we focus on the jet radiative properties rather than its dynamical structure. We consider that only electrons are accelerated up to relativistic energies and produce emissions. However, the cold-proton-matter and magnetic energies dominate dynamics of the jet. The geometry of a low-mass microquasar is the same as a high-mass one. For the latter, the high-mass companion has significant effects in the aspect of radiative output, even on  dynamical structures of the system. However, in the current scenario the companion effect to the system is negligible. The geometry of our jet radiation model is similar to Figure 1 of \citet{Zhang14}, which was developed to study the high-mass microquasar Cygnus X-1.

\subsection{Electron Distribution}
We assume that the steady-state electron distribution in a conical jet is governed by the following kinetic equation \citep[e.g.,][]{ZAA14,Zhang15}
\begin{equation}
\frac{1}{z^2}{\partial \over \partial z} [\Gamma_{\rm j} \beta_{\rm j}cz^2 \tilde{N}(\gamma,z)] + {\partial \over \partial
\gamma} \left[\Gamma_{\rm j} \beta_{\rm j}c\tilde{N}(\gamma,z) {d\gamma \over  dz}\right] = \tilde{Q}_{\rm in}(\gamma,z), \label{dNdz1}
\end{equation}
where the first term represents spatial advection, corresponding to the divergence, $\triangledown\cdot\mathbf{\upsilon} N$, in a spherical coordinate, and the second term energy losses of electrons. $\tilde{Q}_{\rm in}$ is the injection rate of relativistic electrons. $ \tilde{N}$ is the number density of electrons per unit volume, as a function of the electron energy $\gamma$ and the jet height $z$ from the central compact object. $\Gamma_{\rm j}$ is the bulk Lorentz factor of the jet, $\beta_{\rm j}=\sqrt{\Gamma_{\rm j}^2-1}/\Gamma_{\rm j}$ is the bulk velocity, and $c$ is the speed of light. After introducing $N_{\rm \gamma}(\gamma,z)=\Gamma_{\rm j} \beta_{\rm j}cdt^{'}\pi(z{\rm tan}\theta)^2 \tilde{N}$, and $Q_{\rm in}(\gamma,z)=\tilde{Q}_{\rm in}\pi(z{\rm tan}\theta)^2\Gamma_{\rm j} \beta_{\rm j}cdt^{'}$ (here, $\theta$ is a half-cone angle of the jet), Equation (\ref{dNdz1}) is changed into the following form \citep[see also,][]{Moderski03,Zhang14},
\begin{equation}
{\partial N_{\rm \gamma}(\gamma,z) \over \partial z}+{\partial \over \partial
\gamma} \left[N_{\rm \gamma}(\gamma,z) {d\gamma \over dz}\right] = {Q_{\rm in}(\gamma,z)
\over c \beta_{\rm j} \Gamma_{\rm j}}, \label{dNdz2}
\end{equation}
where $Q_{\rm in}=dN_{\rm \gamma}/dt^{'}$. The energy loss rate of relativistic electrons along the jet is given as
\begin{equation}
{d\gamma \over dz} = {1 \over  c\beta_{\rm j} \Gamma_{\rm j}}\left(d\gamma \over dt^{'}\right)_{\rm rad}- {2 \over
3}{\gamma \over z}, \label{dgdz}
\end{equation}
where $t^{'}$ is the time in the jet co-moving frame. The second term of the right-hand side of Equation (\ref{dgdz}) indicates the adiabatic loss of electrons. The total radiative loss rates of an electron, $(d\gamma/dt^{'})_{\rm rad}$, include synchrotron, its Comptonization, and ERC from the disk and corona photons.

It is not clear that which mechanism is operating to accelerate electrons up to relativistic energies. It could be shock acceleration, stochastic acceleration or magnetic reconnection \citep[see][for some discussions]{ZAA14}, or shock interaction in a magnetic reconnection site \citep{Lazarian99,de05}. The diffusive shock acceleration involving the first-order \emph{Fermi} process is often regarded as the most effective scenario in jets. However, this process requires that electrons are pre-heated up the value comparable to the energy of thermal ions. Fortunately, a quasi-Maxwellian electron distribution with strong coupling with protons has been found in PIC simulations of collisionless relativistic \citep{Spitkovsky08,Sironi11} and non-relativistic \citep{Riquelme11} shocks. For the current study, jets are relativistic or should be at leat mildly relativistic. In fact, the low-energy tail of electrons with a thermal distribution has a relatively small contribution to the SEDs. On one hand, a sharp low-energy cutoff, $\gamma_{\rm br}$, from which electrons can be energized, are assumed in \citet{ZAA14} for the jets of black-hole X-ray binaries in the low/hard state. On the other hand, a hard injection spectrum below $\gamma_{\rm br}$, which could approximate the Maxwellian distribution, are used to model a sample of blazars \citep[e.g.,][]{Ghisellini09,Zhangj12}. Similarly to the latter, the relativistic electrons injected in the dissipation region are considered as
\begin{equation}
Q_{\rm in}(\gamma,z)=K_{\rm e} \frac{1}{\gamma^{p}\gamma_{\rm br}^{q-p}+\gamma^{q}},
\end{equation}
where $\gamma_{\rm br}$ is the break energy of relativistic electrons, $p$ and $q$ are the energy spectral indices of the electrons below $\gamma_{\rm br}$ and above $\gamma_{\rm br}$, respectively. The normalization constant of the electrons, $K_{\rm e}$, is determined by
\begin{equation}
K_{\rm e}=\frac{\eta_{\rm rel}\eta_{\rm jet}L_{\rm acc}}{m_{\rm e}c^2\int^{\gamma_{\rm max}}_{\gamma_{\rm min}}\frac{1}{\gamma^{p}\gamma_{\rm br}^{q-p}+\gamma^{q}}\gamma d\gamma}, \label{Lrel}
\end{equation}
where $L_{\rm acc}=\dot{M}_{\rm acc} c^2$ is the accretion power of the system via Roche lobe outflows of the companion star. Here, $\dot{M}_{\rm acc}$ is the mass accretion rate. The power of relativistic electrons, $L_{\rm rel}=\eta_{\rm rel}L_{\rm jet}$, accounts for a fraction of the jet power of $L_{\rm jet}=\eta_{\rm jet}L_{\rm acc}$. We further consider  $\eta_{\rm jet}$ to be 0.1 and $\eta_{\rm rel}$ an adjustable parameter. $\gamma_{\rm min}$ (assuming to be 1) and $\gamma_{\rm max}$ are minimum and maximum energies of the relativistic electron, respectively.

We adopt a theory of shock acceleration to obtain electron maximum energy, considering the balance between an acceleration rate and total loss rates. For a standard first-order \emph{Fermi} acceleration mechanism, the acceleration rate of an electron in an ordered magnetic field $B(z)$ is written as
\begin{equation}
{{\rm d}\gamma \over {\rm d}z}=\frac{\eta_{\rm ac} eB(z)c}{m_{\rm e}c^2}\frac{1}{c\beta_{\rm j}\Gamma_{\rm j}}\ , \label{gain}
\end{equation}
where $\eta_{\rm ac}$ is a parameter that characterizes an acceleration efficiency in a dissipation region and $e$ is the elementary charge. The magnetic fields evolve along the jet in the form of $B(z)=B_{\rm 0}z_{\rm in}/z$.

\subsection{Radiative Processes}
Synchrotron emissions are calculated by using the formulae in \citet{Spada01}. According to the radiative transfer equation, the resulting synchrotron luminosity in solid angle of $\delta \Omega_{\rm j}$ is given by
\begin{equation}
\delta L_{{\rm Syn},\nu'}'(z) = \left[\int \delta j_{\nu'}'(\gamma,\nu')\delta N_{\gamma}(\gamma,z)d\gamma\right]\frac{1-e^{-\tau_{\rm s}}}{\tau_{\rm s}},
\end{equation}
where $\delta j_{\nu'}'$ is the pitch-angle averaged synchrotron emissivity and $\tau_{\rm s}$ is the optical depth of synchrotron self absorption. $\delta N_{\rm \gamma} = N_{\rm \gamma}\delta \Omega_{\rm j}/\Omega_{\rm j}$ is the number density of electrons in each solid angle.

For numerical calculations of both SSC and ERC of disk and corona photons, we used Equation (14) of \citet{Zhang14}. In practice, we consider a spherical corona with a radius of $3R_{\rm in}$ surrounding the central compact object. Here, $R_{\rm in}\approx10R_{\rm g}$ is a truncation radius of the standard thin accretion disk, and  $R_{\rm g}=GM_{\rm BH}/c^2$ is a gravitational radius of the black hole with a mass of $M_{\rm BH}$. It is generally considered that the hard X-ray observations with a power-law spectrum is explained as Comptonization of accretion disk photons by hot electrons in the plasma corona. For simplification, the photon field of the corona is assumed to be a power-law plus an exponential cutoff, $u_{\rm c}=K_{\rm c}\varepsilon^{-p_{\rm c}}e^{-\varepsilon/\varepsilon_{\rm c}}$. Here, $\varepsilon_{\rm c}$ is the maximum energy of the corona photon and the photon index $p_{\rm c}$ is 1.5 in this work. The normalization constant $K_{\rm c}$ are deduced by
\begin{equation}
K_{\rm c}=\frac{L_{\rm cor}}{\int \varepsilon^{-p_{\rm c}}e^{-\varepsilon/\varepsilon_{\rm c}}\varepsilon d\varepsilon},
\end{equation}
where $L_{\rm cor}=f_{\rm c}L_{\rm acc}$ is the bolometric luminosity of the corona and $f_{\rm c}$ is a parameter.

We approximate the accretion disk as a planar geometry following \citet{Sikora13}. The energy density of disk photons in the jet co-moving frame at a height $z$ is expressed as
\begin{equation}
u_{\rm disk}'=\frac{3\Gamma_{\rm j}^2G\dot{M}_{\rm acc}M_{\rm BH}}{4\pi c}\int^{R_{\rm out}}_{R_{\rm in}}\frac{(1-\beta_{\rm j}{\rm cos}\theta_{\rm ex})^2{\rm cos}\theta_{\rm ex}}{(z^2+R^2)R^2}dR,
\end{equation}
where ${\rm cos}\theta_{\rm ex}=z/\sqrt{z^2+R^2}$. $R_{\rm in}$ and $R_{\rm out}$ are the inner and outer radii of the disk, respectively. The emission spectra of an accretion disk
are calculated following \citet{Frank02}, together with an irradiation effect of the outer disk by itself inner zone and/or a corona. Detailed treatment of reprocessing of X-rays in the outer accretion disk would involve many parameters and complex processes; we thus parameterize the reprocessed fraction of bolometric flux with a factor of $f_{\rm out}=0.1$ (see \citealt{Gierlinski09}, for more details). It is noticed that the temperature for a standard thin disk is $T\propto R^{-3/4}$. However, for an irradiation disk, it is $T\propto R^{-1/2}$. Other model details are the same as those of \citet{Zhang14}.

In this study, we consider an extended dissipation region that is located in the jet. In this case, an absorption of high-energy photons can be avoided even if an emission region is very close to the central compact object. As shown in \citet{Vila10} for the low-mass microquasar GX 339--4, the absorption of TeV photons by infrared synchrotron photons is negligible even for a slightly compact dissipation region. From a research experience of high-mass microquasars, we know that at large jet scales absorptions of high-energy photons  by disk and corona photons can be neglected. Therefore, the electromagnetic cascade processes due to the internal synchrotron photon, and the external disk and corona photons are not modeled.

\section{Numerical Results}
\label{Numeri}
In order to test our radiation model, we first explore the evolution of relativistic electrons and multiband SEDs by using some typical parameters for a general low-mass microquasar. For numerical procedures, interested readers are referred to \citet{Zhang14}. The used characteristic parameters for a general low-mass microquasar are given as follows: a mass of black hole of $M_{\rm BH}=10~M_{\rm \odot}$, a distance of $d=2~\rm kpc$, and an accretion rate of $\dot{M}_{\rm acc}=3.0\times10^{-8}M_{\rm \odot}$ yr$^{-1}$. The model parameters are listed in Table \ref{table:cases} for Case A.

\begin{figure*}
  \begin{center}
     \includegraphics[width=120mm,height=100mm,bb=20 210 500 600]{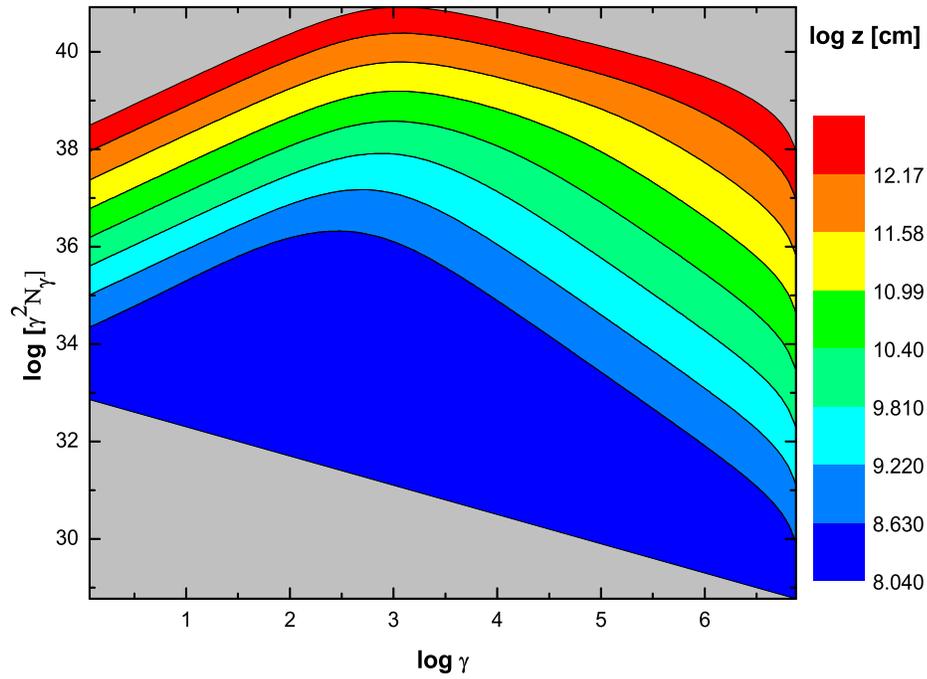}
  \end{center}
\caption{Distributions of relativistic electrons as a function of electron energy at different heights of the jet. The adopted parameters are listed in Table \ref{table:cases} for Case A.}  \label{figs:Elect}
\end{figure*}

\begin{figure*}[]
  \begin{center}
     \includegraphics[width=120mm,height=90mm,bb=20 240 455 600]{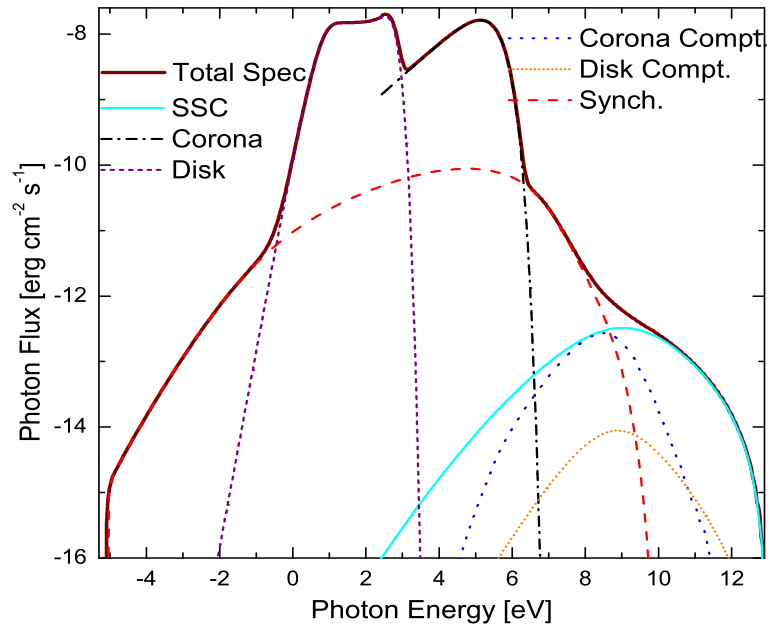}
  \end{center}
\caption{The SEDs of a general low-mass microquasar. The adopted parameters are listed in Table \ref{table:cases} for Case A. The disk spectrum includes the contribution from an irradiated outer disk (short dashed line).}  \label{figs:MODEL}
\end{figure*}

The distributions of relativistic electrons as a function of the energy $\gamma$ along with the height of the jet $z$ are plotted in Figure \ref{figs:Elect}. From the beginning of the injection of relativistic electrons, the number of electrons increases with the height of the jet. It is noticed that when an injection spectrum of electrons is adopted in the form of a broken power law, the resulting electron spectra present the broken form around $\gamma_{\rm b}$. At the low-energy limit, the number of electrons increases with their energies. However, due to a rapidly cooling of electrons, the number of electrons decreases with their energies at the high-energy limit.

Figure \ref{figs:MODEL} demonstrates broadband SEDs for a general low-mass microquasar. As shown in this figure, the emissions from the disk (including a standard thin disk and the corresponding irradiated disk) and hot plasma corona contribute radiative fluxes ranging from near-IR through ultraviolet (UV) to MeV bands. The synchrotron emission provides radiative fluxes at radio, IR, and about $0.1~\rm GeV$ bands. The SSC spectrum dominates fluxes above $1~\rm GeV$. Even though relativistic electrons injected in the jet is at a distance of $70R_{\rm g}$ close to the central compact object, the ERC of soft photons of the accretion disk and corona are still negligible.

\begin{figure*}[]
  \begin{center}
     \includegraphics[width=120mm,height=90mm,bb=20 240 455 600]{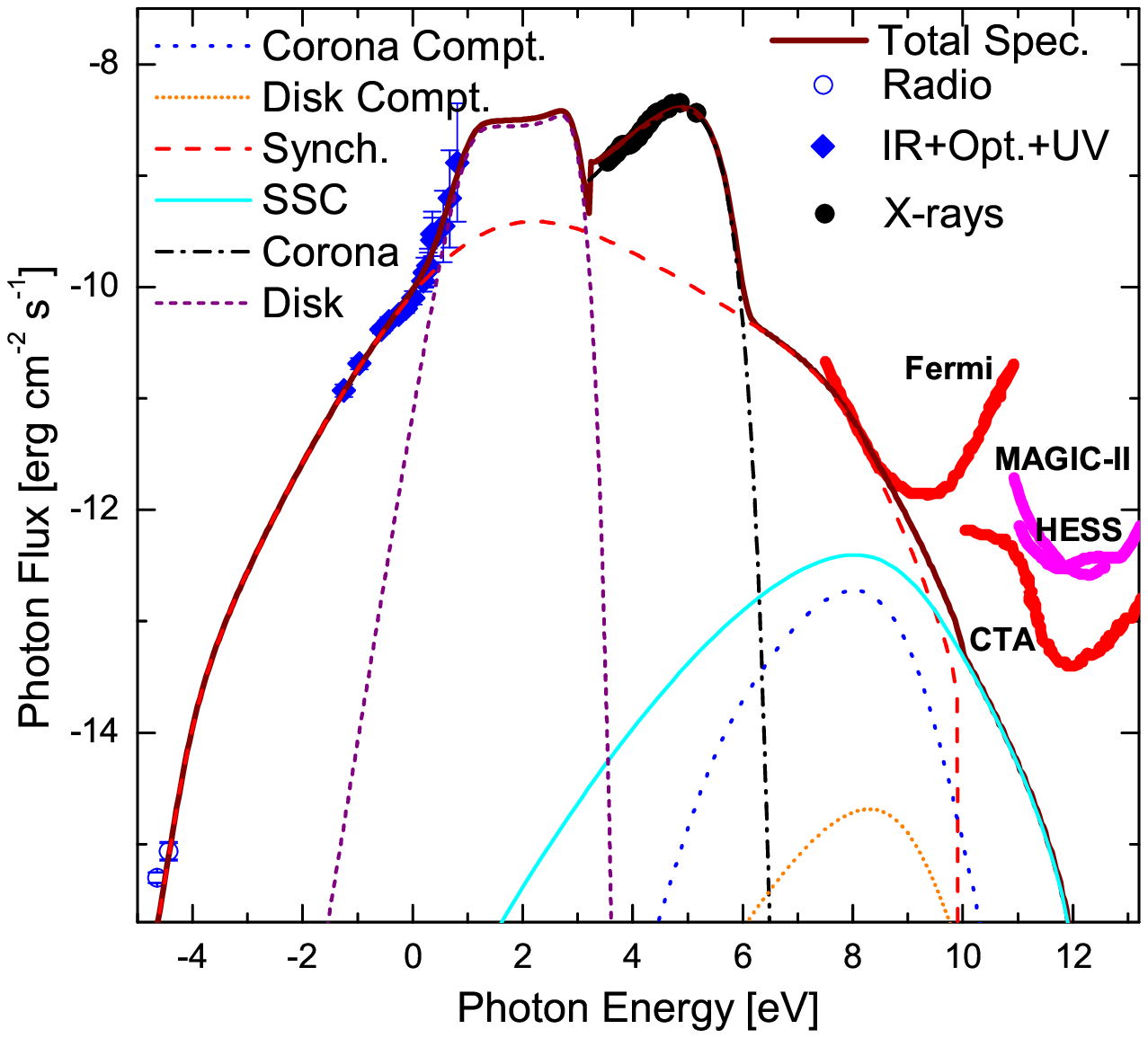}
  \end{center}
\caption{Fitting the SEDs of GX 339--4 for the corona-dominated X-ray emissions under an equipartition condition. The radio through UV up to X-ray observations are from \citet{Gandhi11}. The thick solid lines are $5\sigma$ sensitivity limits for different instruments (1 yr survey mode for \emph{Fermi} and 50 hr of direct exposure for HESS, MAGIC-II and CTA). The fitting parameters are listed in Table \ref{table:cases} for Case B1.}  \label{figs:break-eq}
\end{figure*}

\begin{figure*}[]
  \begin{center}
     \includegraphics[width=120mm,height=90mm,bb=20 240 455 600]{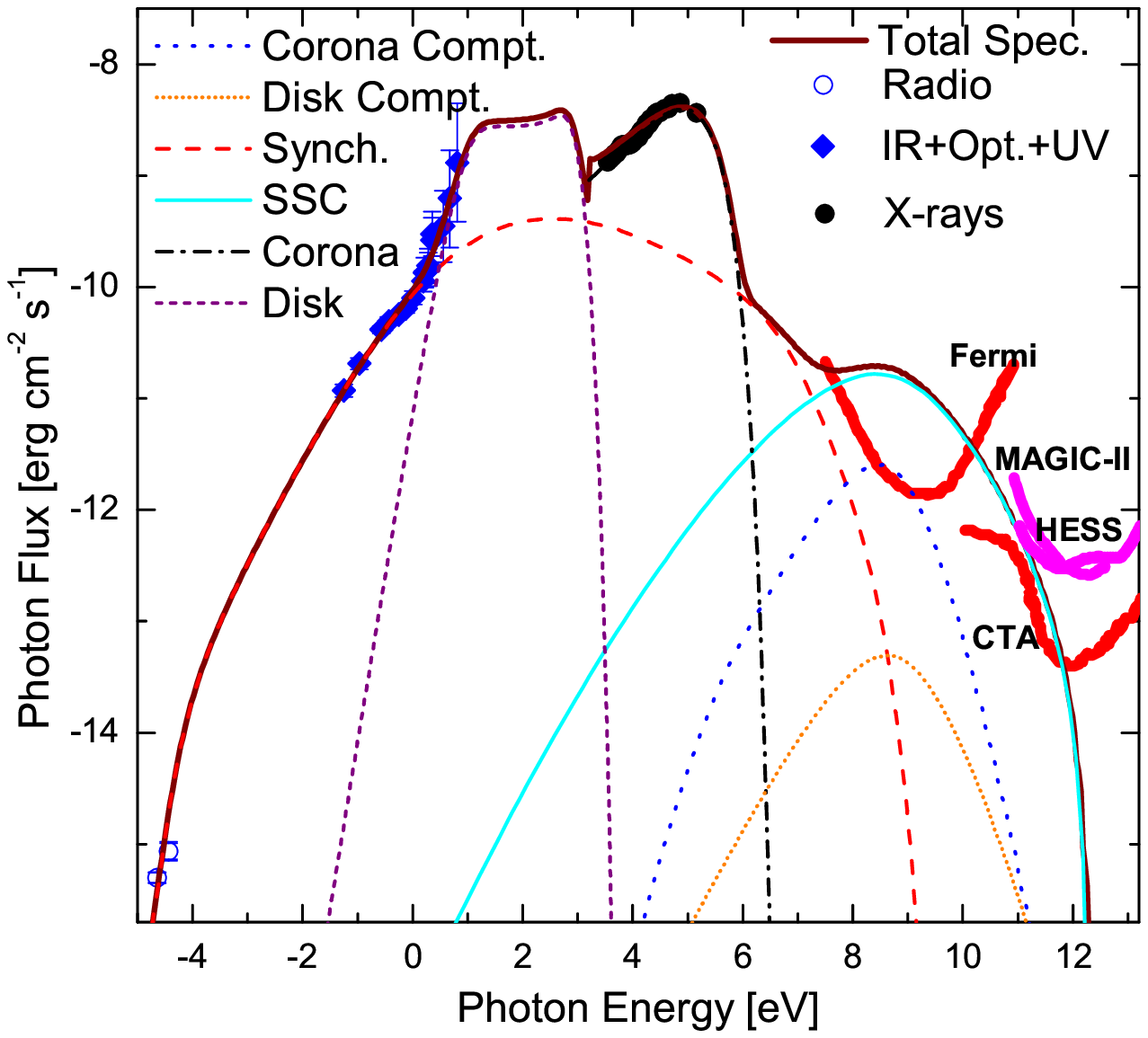}
  \end{center}
\caption{Fitting the SEDs of GX 339--4 for the corona-dominated X-ray emissions under a sub-equipartition scenario. The thick solid lines are the sensitivity limits of \emph{Fermi}, HESS, MAGIC-II and CTA. The fitting parameters are listed in Table \ref{table:cases} for Case B2.}  \label{figs:break-sub}
\end{figure*}

As a result, we find that the nonthermal emissions from the jet, e.g., synchrotron and SSC, dominate the jet radiative outputs. It is well-known that these two components are mainly governed by electron distributions and a magnetic field strength in an emission region. Therefore we define a magnetization parameter $\sigma=u_{\rm B}/u_{\rm kin}$ to characterize a role that magnetic fields play in a jet flow, that is, a ratio of magnetic energy density to matter kinetic density. Here, $u_{\rm B}=B^2/8\pi$ is the magnetic energy density in a completely ordered magnetic field and $u_{\rm kin}=L_{\rm jet}/c\Gamma_{\rm j}\beta_{\rm j}\pi (z{\rm tan}\theta)^2$ is the kinetic energy density of jet matter. With the values adopted in this numerical test, we find $\sigma\sim 4\times10^{-3}$ at the initial location of the dissipation region. It is very evident that the entire dissipation region is sub-equipartition.

\section{Application to GX 339--4}
\label{Appl}
GX 339-4 has been intensively observed in radio, IR, optical, and X-ray
bands \citep[e.g.,][]{Corbel00,Homan05,Yu07,WuYu10,Gandhi11,YanYu12}. The source goes through all the spectral states of a typical black-hole X-ray binary. The radio image of the GX 339--4 jet are achieved by \citet{Gallo04} after the X-ray outbursts of 2002, which confirmed that this source is a microquasar.  A central black hole, with a mass $>6M_{\rm \odot}$, is in an orbit around a dark companion star, with a mass in the range [$0.166M_{\rm \odot}$,$1.1M_{\rm \odot}$] \citep{Munoz08}, which is located at a distance of $\sim 6-9~\rm kpc$ \citep{ZAA04}.

It can be seen that this source is a low-mass microquasar. A distance of $d=6~\rm kpc$ and a black-hole mass of $M_{\rm BH}=6$ are used in our following fittings. In addition, some typical values are fixed as a jet bulk Lorentz factor of $\Gamma_{\rm j}=2$, an accretion rate of $\dot{M}_{\rm acc}=5\times10^{-8}M_{\rm \odot}$ yr$^{-1}$, a half-cone angle of $\theta=5^{\circ}$, and a viewing angle of $\theta_{\rm obs}=30^{\circ}$. As exposed in Section \ref{Numeri}, synchrotron and SSC dominate the radiative outputs of the jet for the case of the sub-equipartition. Below, we are to explore two cases, that is, the magnetization parameter $\sigma\sim1$ and $<<1$.

The first case corresponds to an equipartition between magnetic energy density and kinetic energy density of matter. The initial location of the emission region is first fixed at a distance of $10^8~\rm cm$ ($\sim67.3R_{\rm g}$). We then adjust the values of the jet power $L_{\rm jet}$ and the magnetic field strength $B_{\rm 0}$ to make $\sigma$ close to 1. At last but no least, some parameters that are associated with relativistic electrons, such as spectral indices and break energy, are adjusted to fit observations. According to the constraints of X-ray observations, which are regarded as the corona-dominated emissions, the value of the break energy $\gamma_{\rm br}$ is determined.

Figure \ref{figs:break-eq} shows the SEDs of GX 339--4 confronting with the observations available and the sensitivity limits of telescopes. The fitting parameters are listed in Table \ref{table:cases} for Case B1. As shown in this figure, the synchrotron emission and the irradiation of the accretion disk can well explain observations ranging from radio to UV bands. X-ray data are fitted by the corona photon spectrum that is considered as a power-law plus an exponential cutoff. Synchrotron and SSC spectra are dominant at the GeV and TeV bands, respectively. It can be seen that synchrotron emission flux can compare to the sensitivity limit of \emph{Fermi}, however, SSC flux is significantly below the sensitivity limit of CTA. The ERC spectra from disk and corona are negligible.

\begin{figure*}[]
  \begin{center}
     \includegraphics[width=120mm,height=90mm,bb=20 240 455 600]{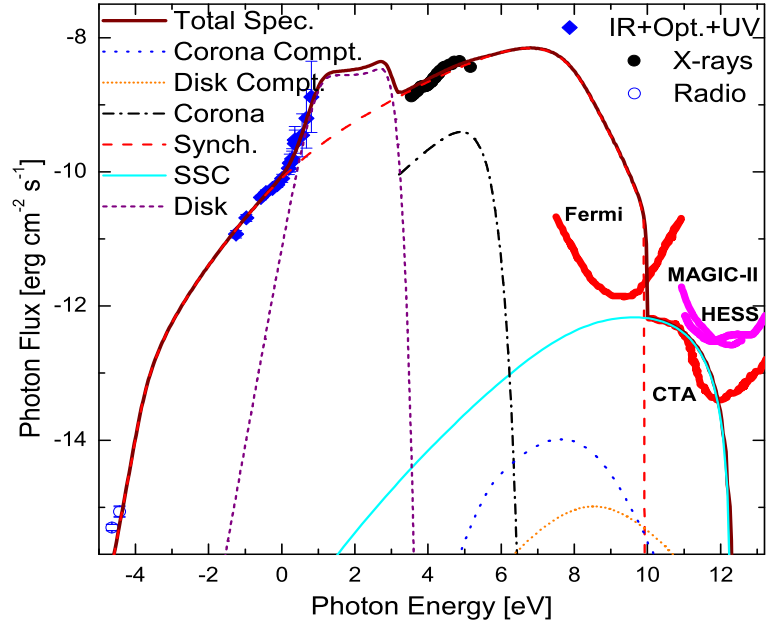}
  \end{center}
\caption{Fitting the SEDs of GX 339--4 for the jet-dominated X-ray emissions under an equipartition condition. The thick lines are the sensitivity limits of \emph{Fermi}, HESS, MAGIC-II and CTA. The fitting parameters are listed in Table \ref{table:cases} for Case C1.}  \label{figs:single-eq}
\end{figure*}

\begin{figure*}[]
  \begin{center}
     \includegraphics[width=120mm,height=90mm,bb=20 240 455 600]{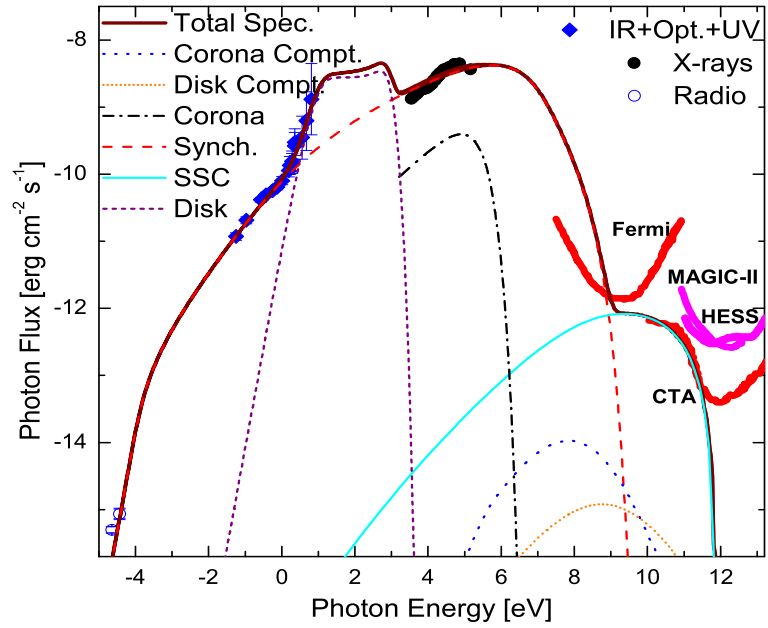}
  \end{center}
\caption{Fitting the SEDs of GX 339--4 for the jet-dominated X-ray emissions under a sub-equipartition scenario. The thick lines are the sensitivity limits of \emph{Fermi}, HESS, MAGIC-II and CTA. The fitting parameters are listed in Table \ref{table:cases} for Case C2.}  \label{figs:single-sub}
\end{figure*}

By adopting the similar method mentioned previously, a sub-equipartition case is explored here. The fitting result is presented in Figure \ref{figs:break-sub} with the parameters given in Table \ref{table:cases} for Case B2. From this figure, we see that at low energy bands, the model can well reproduce observations. SSC component dominates the radiative flux at GeV and TeV bands; this flux can match the sensitivity limits of \emph{Fermi} and CTA, but still below the sensitivity limits of MAGIC-II and HESS. In this case, we obtain a magnetization parameter $\sigma=3.44\times10^{-6}$ at the initial location of electron injection, which indicates that the emission location is in a very low magnetization (e.g., matter-dominated) region.

Hereinbefore, we consider that the X-ray emission has its origin in a corona. However, studies of correlation between X-ray and radio flux show the fact that there is a tight positive correlation for GX 339--4, $F_{\rm radio}\propto F_{\rm X}^{0.7}$ \citep{Corbel03}, which appears to suggest that both waveband emissions have a common origin, that is, they may be from the jet region of X-ray binaries by possible radiative processes  \citep[][see also Section 1 for a review]{Corbel02,Corbel03,Markoff01,Markoff03,Markoff05,Maitra09,Vila10}. After weakening the contribution from the hot corona, we use the jet nonthermal emission to fit X-ray observations together with other band data, considering two scenarios, e.g., equipartition and sub-equipartition.

In order to fit X-ray observations, we have to adjust the spectral index $q$ below 1.5. Meanwhile, we must use a more low break energy $\gamma_{\rm br}$ close to the minimum energy (set as 1) in order to fit IR and optical data. In view of this reason, we give up using the broken power-law spectrum and assume a single power-law form, $Q_{\rm in}\propto\gamma^{-p}$, ranging the entire energy band. Figures \ref{figs:single-eq} and \ref{figs:single-sub} show the fitting SEDs of GX 339--4 for equipartition and sub-equipartition cases, respectively. The used parameters are listed in Table \ref{table:cases} for Case C1 (Figure \ref{figs:single-eq}) and Case C2 (Figure \ref{figs:single-sub}). As shown in these figures, the synchrotron emission can reproduce radiative fluxes at radio, IR, optical, and X-ray bands. SSC component dominates gamma-ray band emissions, which can predict detectable signature by \emph{Fermi} and CTA telescopes, but still below the sensitivity limits of MAGIC-II and HESS. It is noticed that for the case of equipartition the synchrotron emission presents a higher peak frequency than that of the sub-equipartition one ($\sigma=2.39\times10^{-4}$), because a stronger magnetic field is used in the case of the former.

In summary, we have employed our leptonic radiation model to fit observations from GX 339--4 and predict possible gamma-ray emissions. For the two equipartition scenarios (shown in Figures \ref{figs:break-eq} and \ref{figs:single-eq}), it needs to provide a high acceleration efficiency of $\eta_{\rm ac}=0.1$. However, this value is relatively low, that is, 0.01, for the sub-equipartition cases (see also Table \ref{table:cases}). The main reason is that the expected maximum energy of relativistic electrons is required close to about $1~\rm TeV$ in order to emit gamma-rays. In addition, because the dominant (synchrotron) loss rate  is $\propto\gamma^2B^2$ and the acceleration rate $\propto\eta_{\rm ac}B$ (see Equation \ref{gain}), we have the electron maximum energy of $\gamma_{\rm max}\propto(\eta_{\rm ac}/B)^{1/2}$. The fittings show that \emph{Fermi} and CTA should have abilities to detect this source. We note that synchrotron and SSC spectra present different slopes for the different origins of X-ray emissions. Thus, the possible detection with \emph{Fermi} and CTA could help to distinguish the origin of X-ray emissions.

\section{Conclusions and Discussion}
We have proposed a pure leptonic radiation model to study electromagnetic radiation spectra from jets of the low-mass X-ray binaries. This model considers the evolution of relativistic electrons along the jet by numerically solving the kinetic equation. The main photon fields for a general low-mass microquasar, such as accretion disk and corona photons, are involved in the model. Our results show that both synchrotron and SSC spectra play a dominant role, which is similar to the synchrotron plus SSC model that is often used to interpret the emissions of extragalactic BL Lacertae objects. The ERC spectra (disk and corona) are negligible, though the injection location of electrons is close to the central compact object. As an example, we have applied this model to the low-mass microquasar GX 339--4. Our model can well reproduce the observations available at radio, IR, optical, UV, and X-ray bands, and predict potential gamma-ray emissions at the GeV and TeV bands. Observations with instruments, \emph{Fermi} and CTA, can be used to test our model, which could be adopted to distinguish the origin of X-ray emissions.

Based on the work of \citet{Markoff05},  an irradiated disk plus jet model was used to fit observations of the low-mass X-ray binaries, XTE J1118+480 and GX 339--4, from radio through X-rays \citep{Maitra09}. Their results on the fitting of GX 339--4 showed that the jet model emitting synchrotron and SSC radiation dominates emission output throughout the entire waveband, without requiring a disk component, in which SSC component provides a significant fraction of the X-ray emission flux especially when the base of the dissipation region is set to close to the central compact object. In the current work, SSC spectrum dominates the GeV and TeV band emissions and has no contribution to X-ray observations. In addition, an irradiation disk and jet synchrotron or corona components are alternatively adopted to explain emissions from radio to X-ray bands. As shown in Figures 3-6, this work uses more complete IR, Optical and UV data compared with that of Maitra et al. (2009). However, SSC component is still loose due to lacking any data above the X-ray band while the current observations can constrain the synchrotron spectrum.

For the case of the corona-dominated X-ray emissions, locating the dissipation region at sub-equipartition scales is optimistic to gamma-ray productions. Moreover, the value of $\gamma_{\rm br}\sim10^{3}$, which is needed to eliminate the synchrotron spectral contribution to X-ray emissions, is close to the average energy of the shocked-ions. For the case of the jet-dominated X-ray emissions, we used an electron spectrum with a single power-law form. However, in this case, a hard electron spectral index of $p=1.4$ is used. In the theory of non-relativistic diffusive shock acceleration, it is well-known that the spectral index should be close to 2. At present, though there are many works on this topic, we still know little for more physical aspects. We notice that a particle injection spectrum with a hard index has been obtained in \citet{Stecker07}, by using a Monte Carlo simulation of particle diffusive acceleration at relativistic shocks. Besides, formation of hard power-law distribution ($\simeq1$) of the energetic particle spectra is derived by 3D PIC simulations from relativistic magnetic reconnection \citep{Guo14}.

From the perspective of energy spectral fitting, many works regarding the jet emissions indeed demonstrate that the power law index is greater than 2 \citep[e.g.,][]{Markoff05,Kaiser06,Maitra09,Peer09}, in which almost no gamma-ray, especially very high-energy gamma-ray emission is expected. However, if one expects a pure jet model (especially in a low-mass X-ray binary) to produce the entire broadband emissions ranging from radio via optical and X-ray to GeV and TeV bands, the jet itself needs to provide more energy to the energetic electrons. In this case, a hard power-law index is needed, as shown in Figures 5  and 6 \citep[see also][for a similar scenario]{Vila10,Zhang14}, though one would expect a large power law.

As pointed out in Section \ref{figs:MODEL}, an expected location of particle acceleration  should not be at the base of the jet, in which outflow should be a Poynting flux-dominated one. According to the fitting results of GX 339--4, it appears also to support the scenario that the jet emission is away from the central compact object, and at a sub-equipartition region (see also Figures \ref{figs:break-sub} and \ref{figs:single-sub}). In the present work, we parameterized the acceleration efficiency as a factor $\eta_{\rm ac}$. Without a doubt, this overlooks many physical details regarding jets and precludes insight into the jet dynamical properties. Furthermore, in order to obtain the expected maximum energy of relativistic electrons, which mainly decides the condition whether electromagnetic spectra can extend into gamma-ray regimes, we have used the relatively effective efficiency $\eta_{\rm ac}=0.1$ or 0.01.

For the high X-ray luminosities observed in black-hole X-ray binaries or extragalactic blazars, if their emission fluxes are produced in a jet, a powerful jet with a significant amount of leptonic content is necessary, because electrons have very high radiative efficiency in contrast to protons. A lepto-hadronic model needs more effective acceleration efficiency than a pure leptonic scenario, that is, it is necessary to make protons accelerate up to about $10~\rm PeV$, at which protons can efficiently excite  proton--photon/proton--proton interactions. In practice, because the lepto-hadronic model involves multiple secondary particles promoting cascade processes, it is difficult to distinguish the contribution of each individual component. In this regard, our leptonic radiation model is more pure.

\acknowledgments

The authors thank the referee for helpful comments that improved the paper and Dr. Mai-Chang Lei for beneficial discussion. We would like to thank Dr. Poshak Ghandhi for sending us observations. We gratefully acknowledge financial support from the National Basic Research Program of China (973 Program) under grant 2014CB845800, and from the National Natural Science Foundation of China under grants 11233006, 11363003, 11222328, 11333004, 11373002, 11473022 and U1331101.

\begin{deluxetable}{cccccccccccc}
\tabletypesize{}
%\rotate
\tablecaption{The Free Parameters Adopted in the Model.}
\tablewidth{0pt}
\tablehead{
\colhead{Case} & \colhead{$\eta_{\rm rel}$} & \colhead{$z_{\rm in}[R_{\rm g}]$} & \colhead{$z_{\rm end}[R_{\rm g}]$} &  \colhead{$B_{\rm 0}[\rm G]$} & \colhead{$\gamma_{\rm br}$} & \colhead{$p$} & \colhead{$q$} & \colhead{$\eta_{\rm ac}$} & \colhead{$f_{\rm c}$} & \colhead{$\varepsilon_{\rm c}$[$\rm keV$]}}
\startdata
A & 0.1 & $7.00\times10^{1}$ & $3.00\times10^{6}$ & $1.0\times 10^{5}$ & $1.0\times10^{3}$  & 1.0 & 2.5 & 0.01 & 0.01 & 100\\
\tableline
B1 & 0.009 & $6.73\times10^{1}$ & $3.37\times10^{5}$ & $2.0\times 10^{6}$ & $2.0\times10^{3}$  & 1.0 & 2.6 & 0.1 & 0.12 & 60\\
B2 & 0.08 & $6.73\times10^{3}$ & $4.04\times10^{5}$ & $3.0\times 10^{3}$ & $1.0\times10^{3}$  & 1.0 & 2.6 & 0.01 & 0.12 & 60\\
\tableline
C1 & 0.07 & $6.73\times10^{1}$ & $3.37\times10^{5}$ & $3.0\times 10^{6}$ & --  & 1.4 & -- & 0.1 & 0.001 & 60\\
C2 & 0.04 & $6.73\times10^{3}$ & $4.04\times10^{5}$ & $2.5\times 10^{4}$ & --  & 1.4 & -- & 0.01 & 0.001 & 60
\enddata
\tablenotetext{*}{Note. Symbol indicating $\eta_{\rm rel}$: conversion efficiency of relativistic electrons; $z_{\rm in}$: start of dissipation region in jet; $z_{\rm end}$: end of dissipation region in jet; $R_{\rm g}$: gravitational radius; $B_{\rm 0}$: magnetic field strength; $\gamma_{\rm br}$: break energy of electron; $p$: spectral index of electron below $\gamma_{\rm br}$; $q$: spectral index of electron above $\gamma_{\rm br}$; $\eta_{\rm ac}$: acceleration efficiency; $f_{\rm c}$: ratio $L_{\rm cor}/L_{\rm acc}$; $\varepsilon_{\rm c}$: break energy of corona photons.}
\label{table:cases}
\end{deluxetable}

\end{document}